\documentclass[sigconf]{acmart}
\usepackage{multirow}
\usepackage{makecell}
\usepackage{bbding}
\usepackage{diagbox}
\usepackage{mathrsfs}
\usepackage{color}
\usepackage{colortbl}
\usepackage{enumitem}
\usepackage{algorithm}
\usepackage{algpseudocode}
\usepackage{tcolorbox}
\usepackage{balance}
\definecolor{skyblue}{RGB}{135,206,235}

\AtBeginDocument{%
  }

\setcopyright{acmlicensed}
\copyrightyear{2018}
\acmYear{2018}
\acmDOI{XXXXXXX.XXXXXXX}

\acmConference[Conference acronym 'XX]{Make sure to enter the correct
  conference title from your rights confirmation emai}{June 03--05,
  2018}{Woodstock, NY}
\acmISBN{978-1-4503-XXXX-X/18/06}

\copyrightyear{2025}
\acmYear{2025}
\setcopyright{acmlicensed}\acmConference[KDD '25]{Proceedings of the 31st ACM SIGKDD Conference on Knowledge Discovery and Data Mining V.1}{August 3--7, 2025}{Toronto, ON, Canada}
\acmBooktitle{Proceedings of the 31st ACM SIGKDD Conference on Knowledge Discovery and Data Mining V.1 (KDD '25), August 3--7, 2025, Toronto, ON, Canada}
\acmDOI{10.1145/3690624.3709178}
\acmISBN{979-8-4007-1245-6/25/08}
\begin{document}
\arrayrulecolor{black}

\title[Structural Parameterized Adaptation for Efficient Cloud-coordinated On-device Recommendation]{Forward Once for All: Structural Parameterized Adaptation for Efficient Cloud-coordinated On-device Recommendation}

\author{Kairui Fu}
\affiliation{%
  \institution{Zhejiang University}
  \city{Hangzhou}
  \country{China}
}
\email{fukairui.fkr@zju.edu.cn}

\author{Zheqi Lv}
\affiliation{%
  \institution{Zhejiang University}
  \city{Hangzhou}
  \country{China}
}
\email{zheqilv@zju.edu.cn}

\author{Shengyu Zhang}
\authornote{Shengyu Zhang and Kun Kuang are corresponding authors.\label{corresponding}}
\affiliation{%
  \institution{Zhejiang University}
  \city{Hangzhou}
  \country{China}
}
\affiliation{%
  \institution{Shanghai Institute for Advanced Study of Zhejiang University}
  \city{Shanghai}
  \country{China}
}
\email{sy_zhang@zju.edu.cn}

\author{Fan Wu}
\affiliation{%
  \institution{Shanghai Jiao Tong University}
  \city{Shanghai}
  \country{China}
}
\email{fwu@cs.sjtu.edu.cn}

\author{Kun Kuang}
\authornotemark[1]
\affiliation{%
  \institution{Zhejiang University}
  \city{Hangzhou}
  \country{China}
}
\email{kunkuang@zju.edu.cn}

\renewcommand{\shortauthors}{Kairui Fu, Zheqi Lv, Shengyu Zhang, Fan Wu, \& Kun Kuang}

\begin{abstract}
    In cloud-centric recommender system, regular data exchanges between user devices and cloud could potentially elevate bandwidth demands and privacy risks. On-device recommendation emerges as a viable solution by performing reranking locally to alleviate these concerns. Existing methods primarily focus on developing local adaptive parameters, while potentially neglecting the critical role of tailor-made model architecture. Insights from broader research domains suggest that varying data distributions might favor distinct architectures for better fitting. In addition, imposing a uniform model structure across heterogeneous devices may result in risking inefficacy on less capable devices or sub-optimal performance on those with sufficient capabilities. In response to these gaps, our paper introduces \textbf{Forward-OFA}, a novel approach for the dynamic construction of device-specific networks (\textbf{both structure and parameters}). Forward-OFA employs a structure controller to selectively determine whether each block needs to be assembled for a given device. However, during the training of the structure controller, these assembled heterogeneous structures are jointly optimized, where the co-adaption among blocks might encounter gradient conflicts. To mitigate this, Forward-OFA is designed to establish a structure-guided mapping of real-time behaviors to the parameters of assembled networks. 
    Structure-related parameters and parallel components within the mapper prevent each part from receiving heterogeneous gradients from others, thus bypassing the gradient conflicts for coupled optimization. 
    Besides, direct mapping enables Forward-OFA to achieve adaptation through \textbf{only one forward pass}, allowing for swift adaptation to changing interests and eliminating the requirement for on-device backpropagation. Further sophisticated design protects user privacy and makes the consumption of additional modules on device negligible. Experiments on real-world datasets demonstrate the effectiveness and efficiency of Forward-OFA.
\end{abstract}

\begin{CCSXML}
<ccs2012>
<concept>
<concept_id>10002951.10003317.10003347.10003350</concept_id>
<concept_desc>Information systems~Recommender systems</concept_desc>
<concept_significance>500</concept_significance>
</concept>
</ccs2012>
\end{CCSXML}

\ccsdesc[500]{Information systems~Recommender systems}

\keywords{Recommender System, On-device Recommendation}


\maketitle

\section{INTRODUCTION}
\label{intro}
Recent advances in deep learning have significantly enhanced the capabilities of recommender
system, particularly in extracting user preferences from intricate sequential data\cite{zhang2019deep,fu2023end}. Traditional \emph{on-cloud recommendation} methods primarily focus on enhancing the scalability and generalizability of models deployed on cloud. In such systems, user requests are processed on cloud, and recommendation lists are subsequently delivered. This process necessitates transmitting user data between remote devices and cloud, which can introduce substantial network overhead, especially in scenarios characterized by frequent user interactions\cite{yao2022edge,jiang2023ada,zhou2018deep}. Moreover, these user requests often include sensitive information, such as recent item interactions(e.g. \emph{item id}), or even user private profiles(e.g. \emph{ages, income, etc.}). Uploading these sensitive data to cloud may result in a potential leakage of user privacy. 

\begin{figure*}[htb]
    \centering
    \includegraphics[width=1.0\linewidth]{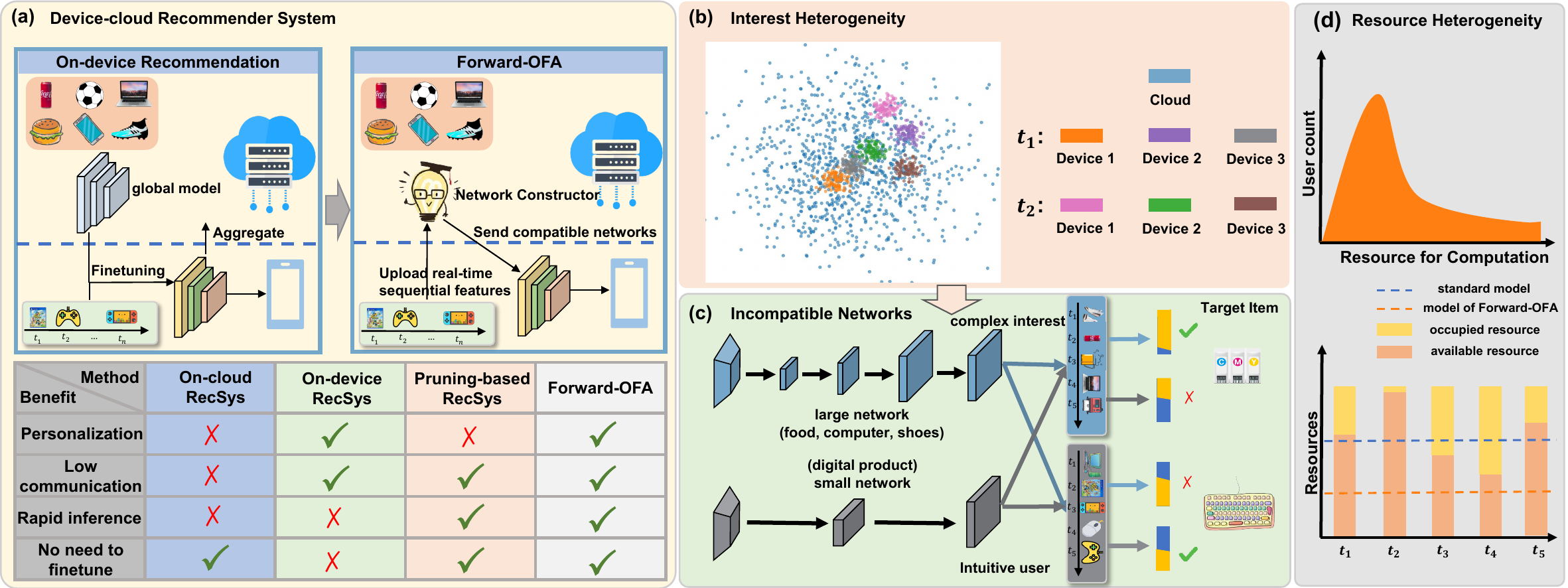}
    \vspace{-0.5cm}
    \caption{(a): Brief comparison between Forward-OFA and other methods used in on-device recommendation. (b): Each device has its own specific behaviors which change frequently while cloud has access to all the historical data of all devices. Distribution shifts among them and within each device make models trained with data on cloud degrade on some devices. (c): Large networks are conducive to exploring complex user interests, while simple networks are suitable enough for intuitive users. (d): Most users own mobile devices that don’t have a lot of computing resources and the computing resources available for the current recommendation task will change in real time due to the presence of other apps.}
    \label{fig: intro}
    \vspace{-0.2cm}
\end{figure*}

Benefiting from the booming computational resources on mobile devices, recommender system is now deploying models directly to mobile devices to better serve users\cite{gong2020edgerec,wu2022fedattack}. This paradigm leverages the computational resources of devices to conduct real-time reranking\cite{gong2020edgerec}, eliminating the need for data uploads to cloud servers for processing. Such an approach not only mitigates network traffic burdens but also significantly enhances user privacy protections. Existing research in this area generally falls into two main categories: personalization-based and cost-aware mechanisms. The former focuses on tailoring device-specific parameters\cite{yao2021device,niu2020billion,liu2021fedct,lv2023duet,lv2024semantic} to enhance the modeling of long-tailed users and items\cite{park2008long}. Contrastingly, cost-aware methods prioritize minimizing both the communication overhead involved in synchronizing local models from cloud and the computational cost of on-device inference due to the requirement for continuous adaptation. For instance, some studies\cite{yao2022device,qian2022intelligent,lv2024intelligent} explore optimal moments for updating models from cloud to prevent unnecessary data transmission. Other works\cite{xia2022device,sreenivasan2022rare,chen2019you,fu2024diet} make an effort to remove redundant parameters for efficient transmission and inference.

While the aforementioned methods have made significant strides in on-device recommendation, their focus has predominantly been on network parameters, potentially overlooking the importance of network structures. Current research demonstrated that the uniqueness of network structures plays a vital role in data distribution modeling\cite{zoph2016neural,zhang2023nasrec}. This phenomenon is illustrated in Figure \ref{fig: intro}(c) of recommender system, where for some intuitive users(the \textcolor{gray}{gray} one) \textcolor{gray}{smaller networks} can already accurately model user interests. In contrast, \textcolor[rgb]{0.25, 0.5, 0.75}{larger networks} are more conducive to modeling complex user interests(the \textcolor[rgb]{0.25, 0.5, 0.75}{blue} one), as these users prefer to explore unknown interests. Figure \ref{fig: intro}(b) reveals the extensive variability of interest between devices themselves and clouds. Consistently deploying identical networks does not ensure satisfactory services for most devices with changing interests. When it comes to \textbf{resource heterogeneity} in Figure \ref{fig: intro}(d), universally applying larger networks will add a substantial burden for devices with limited resources and disrupt the functionality of other applications due to continuous resource usage. On the contrary, applying smaller models for all devices may not fully utilize the capabilities of more robust devices. Consequently, the adaptive construction of networks has become critical for user-oriented recommender systems. Given its demanding and challenging characteristics, this problem raises four crucial research challenges: (\textbf{\romannumeral1}): How to design the corresponding local structures for each device \textbf{accurately}? (\textbf{\romannumeral2}): How to \textbf{efficiently} build adaptive models for a given device to accommodate its local varying interests? (\textbf{\romannumeral3}): How to protect user \textbf{privacy} from exposure to cloud during this process? (\textbf{\romannumeral4}): How to avoid increasing much \textbf{burden} on devices due to the additional modules?

\begin{figure}[h]
  \centering
  % \vspace{-0.2cm}
  \includegraphics[width=0.7\linewidth]{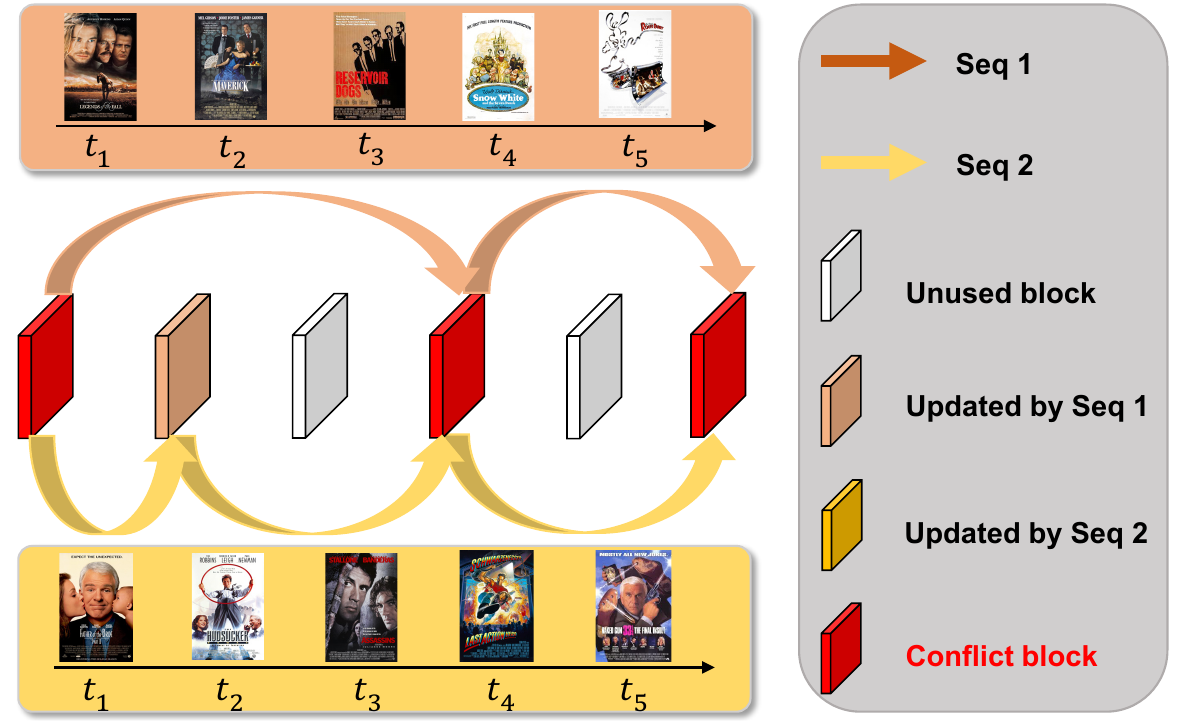}
  \vspace{-0.3cm}
  \caption{When backpropagation, conflict gradients from two sequences with different interests will prevent shared blocks(red blocks) from being updated correctly.}
  \label{fig: conflict}
  \vspace{-0.3cm}
\end{figure}    

In light of this, we propose Forward-OFA(\textbf{Forward} \textbf{O}nce \textbf{F}or \textbf{A}ll), an approach to building both compatible and compact networks for a given device with just one forward pass and without necessitating further on-device backpropagation after training on cloud. Current sequential recommenders, which consist of stacked blocks\cite{kang2018self,tang2018personalized,sun2019bert4rec,yuan2019simple}, guide us to consider learning inference paths containing one or more blocks as specific structures. In particular, we sampled heterogeneous paths from a discrete distribution, which is parameterized by the output of the \emph{structure controller}, to determine each block's existence. 
% 这个地方可能需要说清楚这也是一种weight sharing
As the sample process is discrete and non-differentiable, we utilize the Gumbel Softmax Straight Through Estimator\cite{jang2017cate} to achieve joint optimization of the structure controller and the backbone(\textbf{Challenge \romannumeral1}). However, this feasible approach may encounter gradient conflicts in the co-adaptation among blocks for the coupled optimization of these heterogeneous assembled structures as depicted in Figure \ref{fig: conflict}. 
To navigate this, Forward-OFA aims to learn structure-related parameters for each heterogeneous structure, bypassing the gradient conflicts for coupled optimization. In technique, we propose to learn a structure-guided mapping of real-time behaviors to the parameters of assembled networks, where \emph{the structure-related gradients are directly propagated to the mapper whose components operate individually without heterogeneous gradients from each other}. This structural mapper also allows the construction of adaptive networks in a single forward pass, swiftly adapting to changing interests and eliminating the need for backpropagation on devices. An additional compact constraint ensures that local structures are more deterministic and resource-efficient, addressing \textbf{Challenge \romannumeral2}. To prioritize user privacy, the low-level components of Forward-OFA, responsible for extracting user interest, are stored on device. 
They are designed to be resource-efficient, necessitating far fewer resources compared to traditional recommenders(\textbf{Challenge \romannumeral4}). They only perform inference for adaptive networks when there's a substantial shift of on-device interests\cite{yao2022device,qian2022intelligent}. Instead of raw user information, the extracted latent representation is uploaded to cloud for model construction, thereby protecting user privacy(\textbf{Challenge \romannumeral3}).

Experiments on four real-world datasets underscore the effectiveness and efficiency of Forward-OFA. We conduct several in-depth analyses to uncover the workings of each component within Forward-OFA and the impact of its hyperparameters. Additional network visualization and case studies provide a clear analysis of how Forward-OFA constructs networks and the influence of different networks on diverse devices. In a nutshell, the key contributions of this paper are as follows:
\begin{itemize}
    \item We early attempt to investigate the joint customization of both structure and parameters, analyzing the challenges of interest heterogeneity, network transmission, and on-device inference simultaneously.
    \item We introduce Forward-OFA to adaptively construct compatible networks that strategically remove unnecessary blocks while preserve those compatible ones for each device.
    \item We establish a structural mapping from real-time interactions on each device to its adaptive networks, ensuring both efficient and superior adaptation in a single forward pass.
    \item Extensive experiments demonstrate that Forward-OFA can be applied to various architectures and datasets for better recommendation and a smaller burden.
\end{itemize}

\section{RELATED WORK}

\subsection{Sequential Recommendation}
The field of sequential recommendation has witnessed significant advancements in recent years. Traditional sequential recommenders mainly rely on the Markov assumptions\cite{he2018translation, rendle2010factorizing}, where current interactions only depend on the most recent interaction or interactions, making it challenging to capture long-term dependencies. After the emergence of neural networks\cite{zhang2020devlbert}, neural sequential recommenders have started to utilize advanced techniques to enhance modeling capabilities and account for long-term dependencies\cite{chong2023ct4rec}.  For instance, GRU4Rec\cite{HidasiKBT15}, a solution based on recurrent neural networks, can effectively capture temporal dependencies in sequences. Given multiple history items for each user, determining the importance of them to the target item becomes a critical issue. Attention-based mechanisms\cite{sun2019bert4rec,kang2018self,cen2020controllable,zhang2023adaptive} address it by assigning soft attention weights to items at different time steps, allowing for more flexible integration of information and more precise capture of user interests. Additionally, some CNN-based recommenders\cite{yuan2019simple,tang2018personalized} utilize a convolutional network to process user sequential information by treating the user's embeddings as a picture. However, all the above methods are essentially served on cloud, necessitating frequent message transmission and raising privacy concerns. Compared with them, Forward-OFA and other on-device recommendation methods, are motivated to handle these problems by performing recommendations locally owing to the increasing on-device resources. These model-agnostic frameworks reduce user response time and protect user privacy from exposure to cloud.

\subsection{On-device Recommendation}
Coinciding with the continuous upgrading of computing resources on mobile devices, on-device recommendation is widely used to alleviate network congestion and privacy issues. The majority of current research aims to exploit the diversity of models on devices to enrich personalized user services\cite{xi2023device,chen2021learning}. DCCL\cite{yao2021device} incorporates meta-patch into recommenders to avoid the expensive update of finetuning the entire model, the gradient will be passed to cloud server for aggregation. DUET\cite{lv2023duet} directly sends personalized parameters to device to better adapt to the user's local interest. PEEL\cite{zheng2024personalized}, on the other hand, groups users on cloud and designs a compact and specific item embedding for each group, which not only enhances the personalization but also reduces the space required by embedding. Another method named DIET\cite{fu2024diet} aims at discovering the most compatible parameters for device-specific distribution within a random network to quickly adapt to local changes.  Moreover, to keep pace with the evolving interests of users, recent studies\cite{yao2022device,qian2022intelligent} have introduced methods to detect changes on the device and request new models from cloud\cite{lv2023duet,zheng2024personalized,fu2024diet} when necessary. Nevertheless, current methods mainly focus on parameter adaptation while potentially neglecting the potential of architectures, which is primarily emphasized in our paper.

\section{METHOD}
\begin{figure*}[h]
  \centering
  \includegraphics[width=\linewidth]{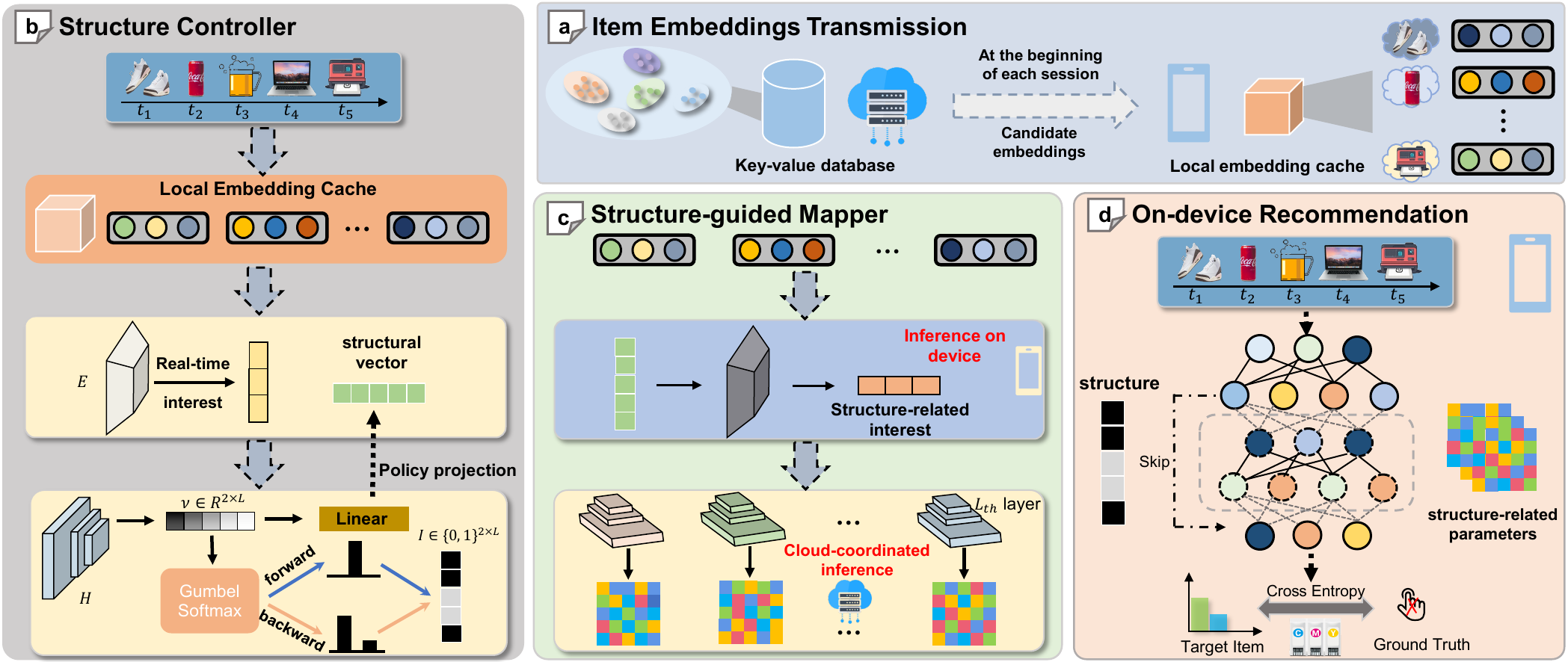}
  \vspace{-0.7cm}
  \caption{Illustrations of all components in Forward-OFA. (a): At the beginning of each session when interest changes dramatically, cloud sends candidate item embeddings to device and those embeddings will be cached for its recommendation in this session. (b): The structure controller consists of an extractor and a lightweight layer for searching the suitable path(distribution vector). The vector will be used in (c) to make structure-related parameters and alleviate the gradient conflict. (c): A mapper to assign personalized and structural parameters, aiming at removing gradient conflicts during training. (d): Each device does not necessarily own the whole network, but only a sub-model to acquire efficiency.}
  \label{fig: method}
  \vspace{-0.4cm}
\end{figure*}
\subsection{Preliminary}
\label{Preliminary}
In recommender system, consider an item set $I=\{i_1, i_2, ..., i_n\}$ and device(user) set $D=\{d_1, d_2, ..., d_m\}$, where $n$ and $m$ represent the number of items and devices. Cloud has the capability to access the historical interactions of each device, denoted as $\mathbb{D}=\{(x_d^t, y_d^t)\}_{d=1,...,m,t=1,...T^d}$, where $T^d$ is the length of the historical interactions of device $d$ and $x_d^t=\{y_d^1,...y_d^{t-1}\}$ denotes all historical interactions before $T^d$. The objective of recommendation is to train a powerful model $M$ using all the data on cloud, which can be formulated as:
\begin{equation}
\label{eq:1}
\begin{aligned}
    \mathop{\min}_{\theta} \sum_{d=1}^{d=m}\sum_{t=1}^{t=T^d}&F(y_d^t|\{y_d^1,y_d^2...,y_d^{t-1}\};\theta), \\  
    &\mathrm{ s.t. } \mathop{\min} \mathbb{P+F}.
\end{aligned}
\end{equation}
Here, $\mathbb{P}$ and $\mathbb{F}$ here refer to the number of parameters and FLOPs for inference of $M$. Besides, each device continue to generate real-time interaction data $X_d=\{y_d^{1^\prime},...,y_d^{{t-1}^\prime}\}$, which may not have appeared on cloud before. Ideally, $M$ should have good generalization ability and show adequate adaptation on these emerging data.

In on-device recommendation, we follow those widely-used paradigms\cite{yao2022device,qian2022intelligent,lv2024intelligent,fu2024diet} from recent works. \emph{At the start of each session, or when user interests change significantly, a series of candidate item embeddings and the updated model are simultaneously sent to devices for local reranking}. These embeddings are then cached on devices in Figure \ref{fig: method}(b). As the recommended items for the user to click are chosen from these candidate items during this session, there is no requirement for any cloud-side involvement during predictions unless local interests change dramatically.

\subsection{Structure Controller for Specific Networks}
\label{Decision Maker}
The use of stacked blocks has become prevalent in recent architectures \cite{kang2018self,yuan2019simple}, significantly advancing the development of recommender systems. However, those numerous blocks also pose a challenge for on-device recommendation, particularly for resource-sensitive devices where a larger network leads to longer computation time. Additionally, complex structures can potentially result in overfitting, especially on devices with limited data. Hence, it is crucial to \emph{precisely allocate appropriate blocks for each device to ensure efficiency and reliability}. A straight approach is to determine where to stop and discard the latter blocks. However, we find that those latter layers might also be crucial for stable prediction as they are closer to the final classifier. In contrast, we tend to predict the presence of each block individually to reserve those later but important blocks. Specifically, we seek a distribution variable $v\in R^{L\times 2}$ for each network, where $L$ is the number of blocks(residual block or transformer block) in a given network. With this variable, for each block $F_l$ with $1\leq l\leq L$, we assemble it if $v_{l,0}\geq v_{l,1}$ otherwise we would skip it. Formally,
% \begin{equation}
%     h_{l+1}=\left\{
%     \begin{aligned}
%         &F_l(h_l) \qquad& \text{argmax}\{v_{l,0}, v_{l,1}\}=0 \\
%         &h_l \qquad& \text{otherwise}
%     \end{aligned}
%     \right.,
% \end{equation}
\begin{equation}
\label{eq:2}
    h_{l+1} = F_l(h_l) * I_{l,0}  + h_l * I_{l,1},
\end{equation}
where $h_l$ and $h_{l+1}$ are the input and output of block $F_l$. Variable $I \in \{0, 1\}^{L\times 2}$ in Equation \ref{eq:2} indicates whether each block $F_l$ is supposed to be executed and its value is determined by $v$, where for $k=\{0,1\}$:
\begin{equation}
\label{eq:3}
    I_{l,k} = \left\{
    \begin{aligned}
        &1 \quad  \text{if} \ \ \text{argmax}\{v_{l,0}, v_{l,1}\} = k \\
        &0 \quad \text{otherwise}
    \end{aligned}
    \right.
\end{equation}

Obviously the gradient of $h_{l+1}$ cannot propagate back to vector $v$ due to the non-differential function $argmax$ in Equation \ref{eq:3}. This inability to optimize through gradient backpropagation leads to the model consistently choosing a single path for parameter optimization, thus failing to achieve the intended purpose of heterogeneous structures. To overcome this, the Gumbel-Softmax straight-through estimator \cite{jang2017cate} is employed to address the non-differentiability of $argmax$. For random variable $G_{i,j}=-log(-log(U_{i,j}))$ sampled from a Gumbel distribution, where $U_{i,j}$ is sampled from a uniform distribution $U(0, 1)$, it is used to reparameterize $v$ to obtain the probability distribution $v^{\prime}$:
\begin{equation}
    v^{\prime}_{i, j} = \frac{e^{(log v_{i,j}+G_{i,j})/\tau}}{\sum_{k=0}^{k=1} e^{(log v_{i,k}+G_{i,k})/\tau}},
\end{equation}
where $\tau$ serves as a temperature control for the smoothness of the sampling process. The closer $\tau$ is to 0, the more the reparameterized variables $v^{\prime}$ resemble the discrete distribution(i.e. one hot representation).  As illustrated in Figure \ref{fig: method}(b), the binary vector $I$ can be sampled during the forward pass, and the gradient of the discrete variable can be approximated by computing the gradient of the continuous softmax relaxation for backpropagation:
\begin{equation}
\label{eq:4}
    I_{i,j} + v^{\prime}_{i, j} - sg(v^{\prime}_{i, j}),
\end{equation}
where $sg$ denotes the stop gradient operation, namely, $sg(x)=x$ and $\nabla sg(x)=0$. 

With Equation \ref{eq:4}, the selected structures can evolve dynamically while trained with the parameters together. However, such a method requires training an individual distribution vector $v$ for each device due to their diverse interests, which is impractical within recommender systems containing substantial and continuously expanding users. The expenses associated with retraining the distribution variable and network parameters for each device can be notably high. Additionally, the scarcity of local data for each device adds to the difficulty of the training process. In response to these difficulties, we propose the direct learning of a mapping from user interactions $X_d$ to its special distribution vector $v_d$:
\begin{equation}
\label{mapping}
X_d = (y_d^{1^\prime},...,y_d^{{t-1}^\prime}) \Longrightarrow v^{2\times L},
\end{equation}
which constructs local preferred sub-structures adaptively based on real-time interests, thereby alleviating the complicated retraining.

As for the mapping process, considering the simplification within Forward-OFA, we utilize a single GRU $E$ to extract user latent interests, followed by a subsequent fully connected layer denoted as $H$, projecting the interest vector onto the space of $v$:
\begin{equation}
\label{beta}
\begin{aligned}
\{\beta_{i,0}, \beta_{i,1}\}_{i=1,...L}&=\{logv_{i,0},logv_{i,1}\}_{i=0,...L}\\
&=H(E(y_d^{1^\prime},...,y_d^{{t-1}^\prime}).
\end{aligned}
\end{equation}

In that case, the structure controller can be trained along with the network parameters. This strategy can also be seen as a way to share knowledge among devices with similar interests\cite{andreas2016neural,kirsch2018modular,sun2020adashare}. In our framework, we place the structure controller on device as it only comprises a single GRU and another fully connected layer. Apart from this, these modules will be utilized to build adaptive sub-structures only when interest changes dramatically or at the beginning of each session\cite{lv2024intelligent,qian2022intelligent}, which momentarily occupies only a few resources.

\subsection{Structural Parameters to Alleviate Conflicts}
\label{Structural Parameters}

Despite the promising approach we developed, there are still underlying challenges when training the structure controller and the original parameters simultaneously. Illustrated in Figure \ref{fig: conflict}, different interactions tend to exhibit varying interests, which consequently requires different network configurations. The optimization of heterogeneous configurations thus introduces another critical problem, \textbf{the co-adaptation of them with the shared parameters} prevents each subnetwork achieves the optimal performance. In Figure \ref{fig: conflict}, the unshared blocks between the two sequences(e.g., the \textcolor{orange}{orange} block owned by Seq 2 in the figure) result in the conflict on both the former block(block 1) and the latter blocks(blocks 4, 5) due to inconsistent gradient directions\cite{bender2018understanding,zhao2021few,hu2022generalizing}: \textbf{\romannumeral1)} The latter blocks receive input from both block 1 and block 2, necessitating a considerable burden to fit both structures sharing the same parameters. \textbf{\romannumeral2)} Regarding the former block, heterogeneous gradients from blocks 2 and 4 are transmitted to it, leading to a more pronounced impact. These discrepancies ultimately result in suboptimal networks.

In contrast to prior approaches utilizing few-shot learning \cite{zhao2021few,hu2022generalizing}, which still encounter potential conflict issues and introduce significant overhead during training, we seek to \textbf{bypass the shared weights when updating the gradients by assigning structure-related parameters for the corresponding networks}(\textbf{conflict \romannumeral1}). Similar to Section \ref{Decision Maker}, given user real-time data $X_d$, Forward-OFA extracts the latent interests in advance and leverages the powerful \emph{hypernetwork}\cite{ha2017hypernetworks} to dynamically generate parameters for each device. As the structures are also determined by device data $X_d$ and its more potential knowledge than the structure variables $\{\beta_{i,0}, \beta_{i,1}\}_{i=0,...L}$, in our experiment it is adopted to construct structure-related parameters to bypass the weight sharing problem.
In technique, only a feature extractor $E^{\prime}$ and $L$ fully connected layers $\{H^{\prime}_k\}_{k=1,..., L}$ are employed to establish a mapping from user real-time data to the parameters of $L$ blocks. 
\begin{equation}
W=\{W_k\}_{k=1,...,L}=\{H^{\prime}_k(E^{\prime}(y_d^{1^\prime},...,y_d^{{t-1}^\prime}))\}_{k=1,...,L}.
\end{equation}
Under these circumstances, gradients would be directly propagated back to $\{E^{\prime}_k\}_{k=1,..., L}$ and $\{H^{\prime}_k\}_{k=1,..., L}$. Moreover, the $L$ layers within the mapper are executed in parallel, thereby there won't be heterogeneous gradients from each other like in Figure \ref{fig: conflict}, overcoming \textbf{conflict \romannumeral2}.

Apart from this, as network structures can also have an impact on the compatible parameters, the output of the structure controller in Section \ref{Decision Maker} is used as the initial state of $E^{\prime}$:
\begin{equation}
\label{eq8}
\begin{aligned}
W&=\{W_k\}_{k=1,...,L}\\
&=\{H^{\prime}_k(E^{\prime}(y_d^{1^\prime},...,y_d^{{t-1}^\prime},\{\beta_{i,0}, \beta_{i,1}\}_{i=0,...L}))\}_{k=1,...,L}.
\end{aligned}
\end{equation}
With structural parameterization, our method constructs specific networks (in both structure and parameters) for each device based on its local interests. Once a device desires to acquire new models, it autonomously builds network paths and sequential characteristics $h=E^{\prime}(y_d^{1^\prime},...,y_d^{{t-1}^\prime},\{\beta_{i,0}, \beta_{i,1}\}_{i=0,...L})$ itself, subsequently uploading them to cloud to acquire compatible and adaptive networks. The whole process is efficient as it only needs one forward inference on cloud without any backpropagation on device.

\subsection{Compact Constraint and Loss Function}
\label{Compact Constraint and Loss Function}
The structural parameterized networks primarily focus on recommendation performance but may overlook the resource constraints of mobile devices. For instance, two sub-structures might achieve comparable performance for a specific user. As the method above does not consider efficiency, it could mistakenly select sub-structures with longer inference time. Towards this end, we introduce another compact constraint $\mathcal{L}_c$ with coefficient $\lambda$ to encourage a lower probability of a block being executed:
\begin{equation}
    \mathcal{L}_c = \sum_{k=1}^{k=L} -log{\alpha_{k,1}},
\end{equation}
where \begin{equation}\alpha_{k,1}=\frac{e^{\beta_{k,1}}}{e^{\beta_{k,0}}+e^{\beta_{k,1}}}.\end{equation}

\begin{table*}[htb]
  \caption{Overall Performance towards both recommendation and resource cost. We use bold font to denote the best-performing model. $\uparrow$ and $\downarrow$ denote that larger and smaller metrics lead to better performance, respectively.} 
\vspace{-0.3cm}
\label{tab:experiment_rec}
\centering
\resizebox{1.0\textwidth}{!}{
    \begin{tabular}{c|c|p{1.15cm}<{\centering}|p{1.15cm}<{\centering}|p{1.15cm}<{\centering}|p{1.15cm}<{\centering}|p{1.15cm}<{\centering}|p{1.15cm}<{\centering}|p{1.15cm}<{\centering}|p{1.15cm}<{\centering}|p{1.15cm}<{\centering}|p{1.15cm}<{\centering}|p{1.15cm}<{\centering}|p{1.15cm}<{\centering}|p{1.15cm}<{\centering}|p{1.15cm}<{\centering}|p{1.15cm}<{\centering}|p{1.15cm}<{\centering}}
    \toprule[2pt]

     \multirow{3}{*}{\textbf{Model}} & \multirow{3}{*}{\textbf{Method}} & \multicolumn{16}{c}{\textbf{Dataset}} \\
     \cline{3-18}
     & & \multicolumn{4}{c|}{\textbf{MovieLens-1M}} & \multicolumn{4}{c|}{\textbf{MovieLens-10M}} & \multicolumn{4}{c|}{\textbf{Amazon-Game}} & \multicolumn{4}{c}{\textbf{Amazon-Food}} \\
     \cline{3-18}
     & & \textbf{NDCG $\uparrow$} & \textbf{Hit $\uparrow$} & \textbf{FLOPs $\downarrow$} & \textbf{Param $\downarrow$} & \textbf{NDCG $\uparrow$} & \textbf{Hit $\uparrow$} & \textbf{FLOPs $\downarrow$} & \textbf{Param$\downarrow$} & \textbf{NDCG $\uparrow$} & \textbf{Hit $\uparrow$} & \textbf{FLOPs $\downarrow$} & \textbf{Param $\downarrow$} & \textbf{NDCG $\uparrow$} & \textbf{Hit $\uparrow$} & \textbf{FLOPs $\downarrow$} & \textbf{Param $\downarrow$} \\
     \bottomrule[1pt]
     \bottomrule[1pt]

     \multirow{8}{*}{\textbf{SASRec}} & \texttt{DeviceRec} & 0.0969 & 0.1816 & 0.6244 & 3.9936 & 0.0718 & 0.1308 & 0.6244 & 3.9936 & 0.0359 & 0.0541 & 0.6244 & 3.9936 & 0.0526 & 0.0620 & 0.6244 & 3.9936 \\
     & \texttt{Finetune} & 0.0939 & 0.1793 & 0.6244 & 3.9936 & 0.0681 & 0.1272 & 0.6244 & 3.9936 & 0.0343 & 0.0537 & 0.6244 & 3.9936 & 0.0523 & 0.0623 & 0.6244 & 3.9936  \\
     & \texttt{DUET}  & 0.1140 & 0.2050 & 0.6244 & 3.9936 & 0.0871 & 0.1511 & 0.6244 & 3.9936 & 0.0456 & 0.0676 & 0.6244 & 3.9936 & 0.0582 & 0.0701 & 0.6244 & 3.9936  \\
     & \texttt{STTD} & 0.0974 & 0.1823 & 3.5735 & 3.7478 & 0.0718 & 0.1314 & 3.5735 & 3.7478 & 0.0371 & 0.0588 & 3.5735 & 3.7478 & 0.0541 & 0.0648 & 3.5735 & 3.7478 \\
     & \texttt{Rare Gem} & 0.0970 & 0.1838 & 0.6244 & 3.9936 & 0.0741 & 0.1344 & 0.6244 & 3.9936 & 0.0373 & 0.0577 & 0.6244 & 3.7696 & 0.0533 & 0.0635 & 0.6244 & 3.6364 \\
     & \texttt{Gater} & 0.0980 & 0.1821 & 0.5738 & 3.9936 & 0.0733 & 0.1398 & 0.5863 & 3.9936 & 0.0393 & 0.0600 & 0.5858 & 3.9936 & 0.0535 & 0.0644 & 0.6050 & 3.9936 \\
     \cline{2-18}
     \rowcolor{gray!40}
     \cellcolor{white} & \texttt{Forward-OFA} & \textbf{0.1182} & \textbf{0.2081} & \textbf{0.5699} & \textbf{3.6384} & \textbf{0.0905} & \textbf{0.1566} & \textbf{0.4795} & \textbf{3.0522} & \textbf{0.0487} & \textbf{0.0754} & \textbf{0.5720} & \textbf{3.7216} & \textbf{0.0590} & \textbf{0.0713} & \textbf{0.5667} & \textbf{3.6160}  \\
     \rowcolor{skyblue!70}
     \cellcolor{white} & \texttt{Improv.$\uparrow$} & 21.97$\%$ & 14.59$\%$ & $\times$ 1.10 & $\times$ 1.10 & 26.03$\%$ & 19.68$\%$ & $\times$ 1.30 & $\times$ 1.31 & 35.61$\%$ & 39.28$\%$ & $\times$ 1.09 & $\times$ 1.07 & 12.29$\%$ & 14.90$\%$ & $\times$ 1.10 & $\times$ 1.10  \\
    
    \midrule[1pt]
    \midrule[1pt]

    \multirow{8}{*}{\textbf{NextItNet}} & \texttt{DeviceRec} & 0.0975 & 0.1846 & 1.5053 & 9.5846 & 0.0638 & 0.1166 & 1.5053 & 9.5846 & 0.0275 & 0.0440 & 1.5053 & 9.5846 & 0.0401 & 0.0467 & 1.5053 & 9.5846  \\
    & \texttt{Finetune} & 0.0879 & 0.1715 & 1.5053 & 9.5846 & 0.0635 & 0.1165 & 1.5053 & 9.5846 & 0.0279 & 0.0422 & 1.5053 & 9.5846 & 0.0399 & 0.0467 & 1.5053 & 9.5846  \\
     & \texttt{DUET} & 0.1175 & 0.2058 & 1.5053 & 9.5846 & 0.0815 & 0.1369 & 1.5053 & 9.5846 & 0.0448 & 0.0656 & 1.5053 & 9.5846 & 0.0515 & 0.0606 & 1.5053 & 9.5846  \\
    & \texttt{STTD} & 0.0920 & 0.1725 & 4.1595 & 4.5957 & 0.0612 & 0.1108 & 4.1595 & 4.5957 & 0.0254 & 0.0421 & 4.1595 & 4.5957 & 0.0394 & 0.0454 & 4.1595 & 4.5957 \\
    & \texttt{Rare Gem} & 0.1035 & 0.1921 & 1.5053 & 5.4784 & 0.0524 & 0.0998 & 1.5053 & 5.4784 & 0.0328 & 0.0496 & 1.5053 & 5.4784 & 0.0417 & 0.0477 & 1.5053 & 5.4784 \\
    & \texttt{Gater} & 0.0981 & 0.1849 & 0.7141 & 3.8656 & 0.0613 & 0.1119 & 0.6107 & 7.7760 & 0.0304 & 0.0426 & 0.6924 & 8.8224 & 0.0385 & 0.0438 & 0.5718 & 7.2832 \\
    \cline{2-18}
     \rowcolor{gray!40}
     \cellcolor{white} & \texttt{Forward-OFA} & \textbf{0.1226} & \textbf{0.2140} & \textbf{0.6167} & \textbf{3.8656} & \textbf{0.0826} & \textbf{0.1429} & \textbf{0.4324} & \textbf{2.6816} & \textbf{0.0477} & \textbf{0.0708} & \textbf{0.6228} & \textbf{3.9072} & \textbf{0.0523} & \textbf{0.0613} & \textbf{0.3126} & \textbf{1.9104} \\
     \rowcolor{skyblue!70}
     \cellcolor{white} & \texttt{Improv.$\uparrow$} & 25.74$\%$ & 15.96$\%$ & $\times$2.44 & $\times$2.48 & 29.36$\%$ & 22.60$\%$ & $\times$3.48 & $\times$3.57 & 73.39$\%$ & 66.67$\%$ & $\times$2.42 & $\times$2.45 & 30.24$\%$ & 30.99$\%$ & $\times$4.82 & $\times$5.02 \\

    \bottomrule[2pt]
    \end{tabular}
}
\end{table*}

In addition to reducing model capacity and decreasing on-device response time, $\mathcal{L}_c$ can assist Forward-OFA in identifying the most effective blocks(increasing their scores) and discarding less effective blocks, thus enhancing the performance of the framework.

Finally, the overall loss function $\mathcal{L}$ is defined as:
\begin{equation}
    \mathcal{L} = \mathcal{L}_{rec} + \lambda \mathcal{L}_c,
\end{equation}
where $\mathcal{L}_{rec}$ denotes the cross-entropy loss defined in Equation \ref{eq:1} between the prediction and the ground truth(next-clicked item).  \textbf{A detailed pseudocode is used in the Appendix \ref{Pseudocode} to clarify the workflow of Forward-OFA better.}

% \begin{equation}
% \begin{aligned}
%     \mathcal{L} &= \mathcal{L}_{rec} + \lambda \mathcal{L}_c \\
%     &= dist(y_d^t, \mathcal{M}(x_d^t, E, H, E^{\prime}, H^{\prime}, W)) \\
%     &\,\,\,+ \lambda \sum_{k=1}^{k=L} -\alpha_{l,1} log{\alpha_{l,1}}
% \end{aligned}
% \end{equation}

\section{EXPERIMENT}

In this section, we conduct various experiments on two widely used sequential recommenders using four real-world datasets, aiming to address the following research questions:

\begin{itemize}[leftmargin=*]
\item \textbf{RQ1:} How does Forward-OFA perform compared to other on-device recommendation methods in terms of resource consumption and recommendation?
\item \textbf{RQ2:} How do the proposed modules of Forward-OFA and different hyperparameters affect the final performance?
\item \textbf{RQ3:} Do the additional modules for adaptive network construction in Forward-OFA impose significant burdens on devices?
\item \textbf{RQ4:} What impact does an adaptive network have on different user distributions?
\end{itemize}

\subsection{Experimental Setup}

\subsubsection{Datasets.} We adopt Movielens-1M, MovieLens-10M\cite{movielens}, Amazon-Game, and Amazon-food\cite{ni2019justifying}, four real-world datasets in the experiment. The data preprocessing can be found in Appendix \ref{preprocess}.

\subsubsection{Evaluation metrics.} Through our experiments, we employ the widely adopted metrics\cite{zhang2021causerec,li2023text} NDCG and Hit in our experiments to assess the accuracy of recommendations provided by each method for devices. In terms of resource constraints, we primarily focus on the inference time on devices, calculated using FLOPs (floating point operations) of local networks during a single inference. Additionally, users are required to update the latest models from the cloud as needed to meet their requirements. Therefore, the size of network transmission, which is reflected by the number of bits in the recommenders(\emph{param}), should also be taken into consideration. It is important to note that \textbf{higher NDCG and Hit mean better performance while smaller FLOPs and params indicate less dependence on resources}. Consistently, we utilize NDCG@10 and Hit@10 throughout the experiment for uniformity.

\subsubsection{Baseline.} To demonstrate the effectiveness of Forward-OFA, we adopt two widely used sequential recommender \emph{SASRec}\cite{kang2018self} and \emph{NextItNet}\cite{yuan2019simple} as base models, as they are made up of transformer and convolutional blocks, separately. We select 6 methods(\emph{Device-Rec}, \emph{Finetune}, \emph{DUET}\cite{lv2023duet}, \emph{Rare Gem}\cite{sreenivasan2022rare}, \emph{STTD}\cite{xia2022device} and \emph{Gater}\cite{chen2019you}) as baselines. \textbf{More descriptions can be found in the Appendix.}

\subsubsection{Implementation Details}

The SASRec and NextItNet we used in the experiment are 6 and 12 blocks, respectively. We follow consistent training epochs for all models, except for \emph{Finetune} and \emph{Rare Gem}, which require additional on-device tuning and sparse network retraining. The learning rate is set to 0.001 for \emph{SASRec} and 0.002 for NextItNet, respectively. Specifically for \emph{SASRec}, the coefficient for resource constraint $\lambda$ is 0.005 for MovieLens-1M and Amazon-food, 0.001 for Amazon-game, and 0.01 for MovieLens-10M. For NextItNet, all datasets adopt 0.001 as the coefficient. In \emph{SASRec}, the temperature $\tau$ is set to 5, 6, 5, and 7 for MovieLens-1M, MovieLens-10M, Amazon-game, and Amazon-food, respectively. In \emph{NextItNet}, the temperature values are set to 12, 7, 6, and 6 for the corresponding datasets. To ensure a fair comparison, \emph{we have adjusted the hyperparameters of compression-based methods and compared their performance with Forward-OFA under the same device constraints}.

\subsection{Overall Analysis(RQ1)}

To demonstrate the potential of Forward-OFA, we conduct experiments on four real-world benchmarks, comparing them to a range of baselines. The results are shown in Table \ref{tab:experiment_rec}. From these results, we can conclude that although deploying models on devices reduces the frequency of transmission when requesting, \emph{DeviceRec} demands the same number of floating-point operations for inference, which is not practical for most devices with limited computing resources. Additionally, consistent networks for all devices may produce ambiguous recommendations for some users compared to those with adaptive networks, owing to the distinct distribution between them and cloud. It also requires enormous bandwidth to request the latest models from cloud in Figure \ref{intro}(a).

\emph{STTD} and \emph{Rare Gem} aim to compress the model parameters to reduce the transmission cost during the cloud-coordinated model updating process. However, neither of them effectively mitigates the overhead caused by on-device inference. In fact, \emph{STTD} even increases the computational cost due to the additional computation required for the semi-tensor product’s matrix multiplication. Additionally, the random distribution of 0s in the sparse matrix obtained by \emph{Rare Gem} still necessitates calculations once there are non-zero elements in each layer. \emph{Gater} speeds up local inference but fails to achieve a satisfactory compression ratio without ensuring comparable performance. Furthermore, these methods still exhibit degradation on certain datasets, notably in Movielens-10M of NextItNet, where \emph{Rare Gem} reduced the NDCG and Hit by 16$\%$.

\emph{Finetune} leverages limited interactions on devices to adjust the parameters. However, as depicted in Table \ref{tab:experiment_rec}, it does not yield provide satisfying results compared to \emph{DeviceRec}. We attribute this phenomenon to the overfitting problem for devices with few interactions and changing interest between training and test data. Another personalized method \emph{DUET} achieves relatively large improvements in both NDCG and Hit. Personalized parameters for each user can effectively capture the latent interests of each user in the behavior sequence. Nevertheless, as mentioned in Section \ref{intro}, the consistent model structures in \emph{DUET} may not be effective enough to serve some users. The same inference overhead as the original model also limits its practical application.

On the contrary, \emph{Forward-OFA} constructs both local adaptive parameters and structures for each device and adopts a structural mapper to facilitate efficient adaptation. This approach not only achieves satisfying performance but also minimizes resource consumption, encompassing both transmission delay and inference. Notably, \emph{Forward-OFA} outperforms other baselines on all four metrics, underscoring the effectiveness of our framework. In particular, \emph{Forward-OFA} improved by nearly 20$\%$ on the Movielens-1M and Movielens-10M datasets, and by 30$\%$ on Amazon-food of \emph{NextItNet}. It’s worth noting that the \emph{NextItNet} utilized in our experiments has more blocks than \emph{SASRec} (6 vs 12), consequently resulting in relatively smaller FLOPs on \emph{NextItNet} compared to those on \emph{SASRec}.

\subsection{In-depth Analysis(RQ2)}

\subsubsection{Block visualization}
To gain a deeper understanding of the impact of our framework, we conduct visualizations of block distributions and experiments to explore the influence of each component on adaptive networks. In Figure \ref{fig: vis}, we plot the distribution of local adaptive networks and usage of blocks in the test set, which can provide insight into the functioning of our method. Based on the figure, we have the following observations: 1) Different blocks exhibit significant variations in the number of times they are selected. Some blocks are infrequently chosen, while others are deemed more crucial for most users, reflecting the diverse interests of users on the device. Most users tend to include the first and last blocks in their local networks as the former mainly extract low-level knowledge of user interactions and the latter is responsible for the final sequential representation in recommender system. 2) Not all users require all blocks within their local network. On the contrary, only a small fraction of users tend to retain the entire network, demonstrating the significance of building adaptive networks for each user.

\subsubsection{Influence of adaptive networks}
\begin{figure}[h]
  \centering
  \includegraphics[width=\linewidth]{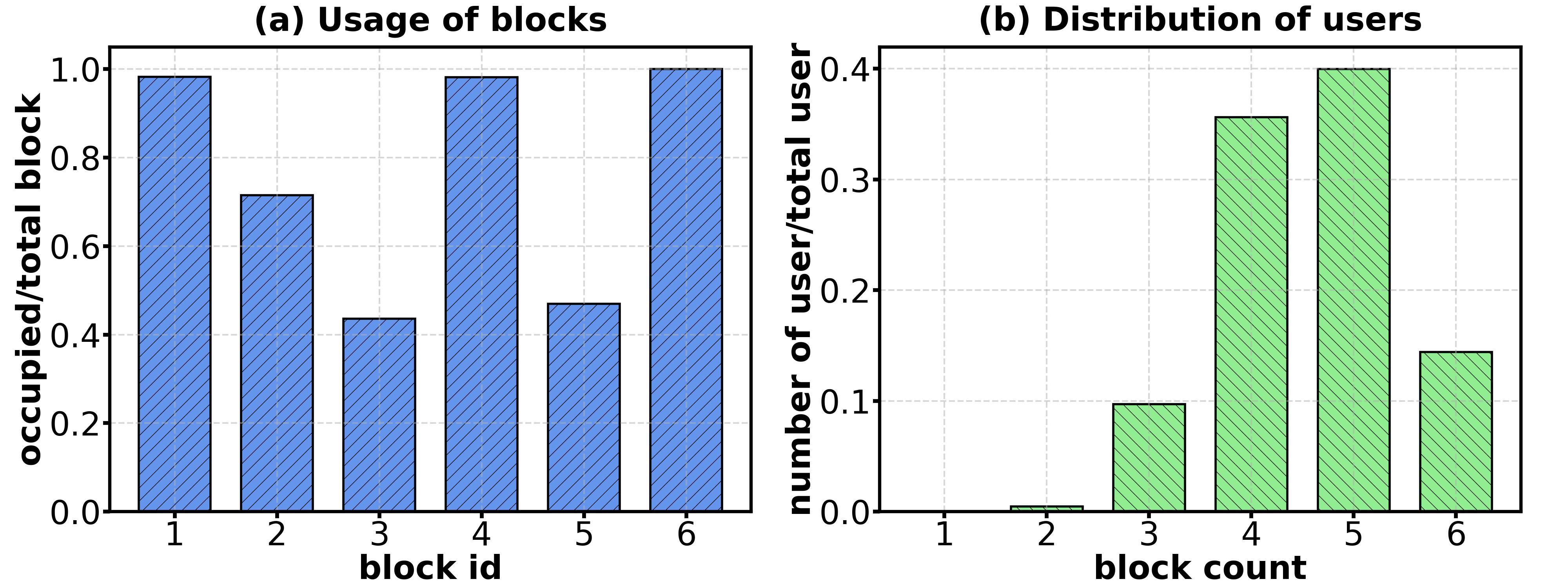}
  \vspace{-0.3cm}
  \caption{Analysis conducted on Movielens-10M using a 6-layer SASRec. (a) The number of times each block is selected by the device in the test set. (b) Distribution of users with different numbers of blocks.}
  \label{fig: vis}
  \vspace{-0.2cm}
\end{figure}

\begin{table}[htb]
\caption{Analysis of the structure controller and structural mapper involved in constructing adaptive networks.} 
\vspace{-0.2cm}
\label{tab: controller}
\centering
\resizebox{1.0\linewidth}{!}{
    \begin{tabular}{c|c|c|c|c|c|c|c}
    \toprule[2pt]

     \multirow{3}{*}{\textbf{Model}} & \multirow{3}{*}{\textbf{Method}} &  \multicolumn{6}{c}{\textbf{Dataset}} \\
     \cline{3-8}
     & & \multicolumn{2}{c|}{\textbf{MovieLens-1M}} & \multicolumn{2}{c|}{\textbf{MovieLens-10M}} & \multicolumn{2}{c}{\textbf{Amazon-Game}} \\
     \cline{3-8}
     & & \textbf{NDCG} & \textbf{Hit} & \textbf{NDCG} & \textbf{Hit} & \textbf{NDCG} & \textbf{Hit} \\
     \bottomrule[1pt]
     \bottomrule[1pt]
     
     \multirow{3}{*}{\textbf{SASRec}} & \texttt{DeviceRec} & 0.0969 & 0.1816 & 0.0718 & 0.1308 & 0.0359 & 0.0541  \\
     & \texttt{+controller} & 0.0946 & 0.1765 & 0.0703 & 0.1289 & 0.0369 & 0.0573   \\
     & \texttt{+mapper} & 0.1142 & 0.2043 & 0.0868 & 0.1505 & 0.0452 & 0.0678 \\

     \midrule[1pt]
     \midrule[1pt]

    \multirow{3}{*}{\textbf{NextItNet}} & \texttt{DeviceRec} & 0.0975 & 0.1846 & 0.0638 & 0.1166 & 0.0275 & 0.0425 \\
     & \texttt{+controller} & 0.0973 & 0.1841 & 0.6114 & 0.1117 & 0.0316  & 0.0458  \\
     & \texttt{+mapper} & 0.1193 & 0.2088 & 0.0816 & 0.1422 & 0.0461 & 0.0692  \\

    \bottomrule[2pt]
    \end{tabular}
}
\vspace{-0.2cm}
\end{table}

Table \ref{tab: controller} presents the performance of the ablation models and the base recommender \emph{DeviceRec}. \emph{+controller} denotes the augmentation of the base model with a structural controller to predict the network structure for devices. From the table, we can conclude that in this way this approach does not always lead to satisfactory performance, despite the compact model it selects as explained in Figure \ref{fig: vis}. As introduced in Section \ref{Structural Parameters}, gradient conflicts result in incorrect updates within the shared blocks. With fewer blocks in \emph{SASRec}, more users opt to share blocks, making it more pronounced in such scenarios. However, we can still observe some improvements on Amazon-Game where the conflict problem is relatively small and \emph{+controller} consistently performs better than the base recommender. This finding demonstrates heterogeneous structures are indeed beneficial for a precise recommendation. On the other hand, \emph{+mapper} bypasses the updates of shared blocks by associating parameters with user interests and the network structure, thus significantly enhancing the model’s effectiveness.

\subsubsection{Influence of sparsity constraint}
\begin{figure}[h]
  \centering
  \includegraphics[width=\linewidth]{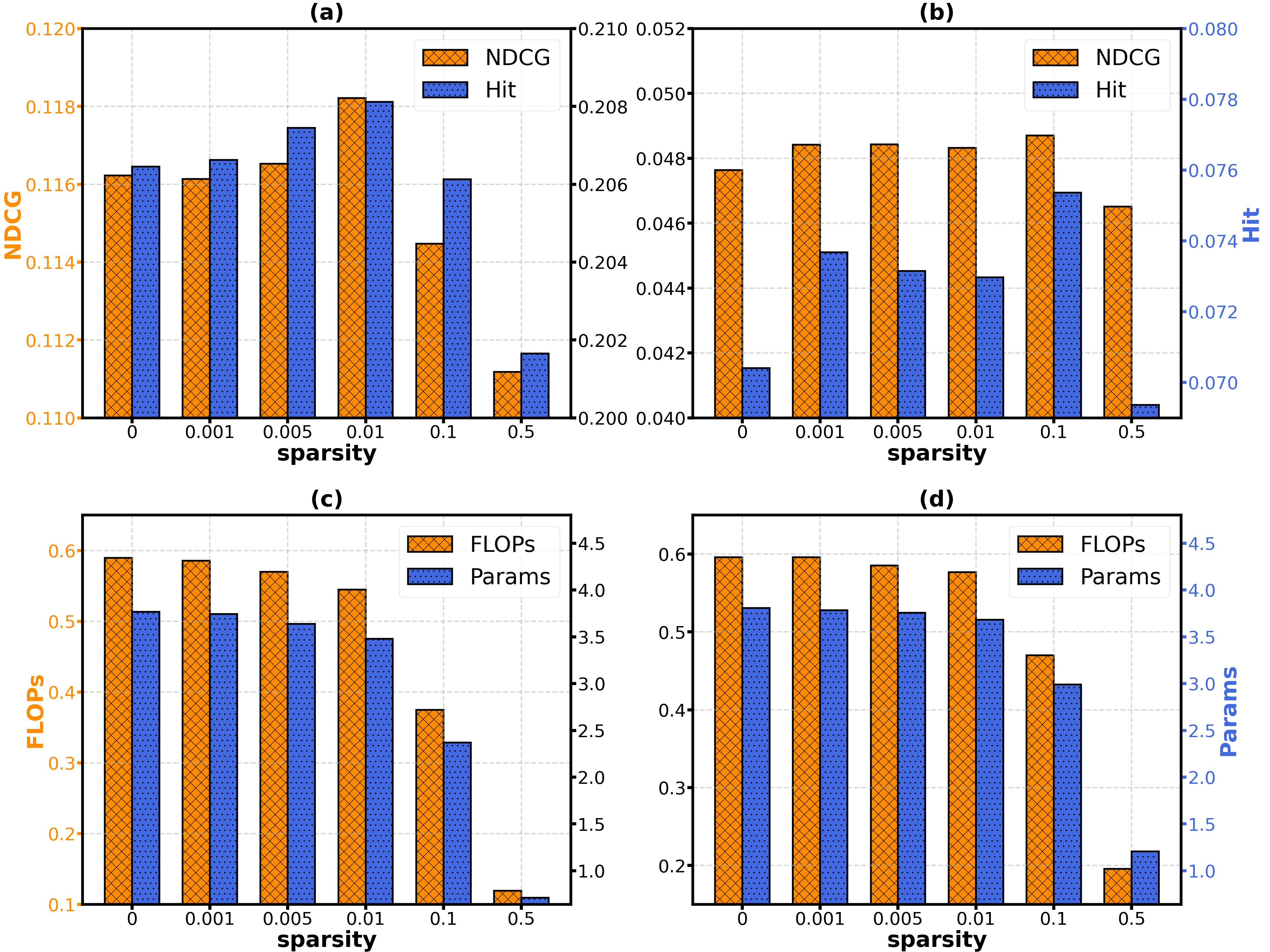}
  \vspace{-0.5cm}
  \caption{The influence of the coefficient $\lambda$ of the sparsity constraint. The first and second columns are the experimental results on Movielens and Amazon-game respectively. (a) and (b) report the results on NDCG and Hit while (c) and (d) represent the results on FLOPs and params.}
  \label{fig: sparsity}
  \vspace{-0.4cm}
\end{figure}

We are interested in understanding how the sparsity constraint $\mathcal{L}_c$ in Section \ref{Compact Constraint and Loss Function} influence our method, therefore we adjust the coefficient and plot the result in Figure \ref{fig: sparsity}. First, when we increase the coefficient $\lambda$, the average FLOPs and params in the test set will continue to decline as higher $\lambda$ forces the probability of unnecessary blocks being selected to decrease. Furthermore, even without the sparsity constraint, the local models still require fewer resources than the original recommender in Table \ref{tab:experiment_rec}, thus proving the effectiveness of our structure controller. Second, an appropriate constraint could help our choice to be more deterministic and trust those more effective blocks, thereby enabling the learning of better subnetworks. Figures \ref{fig: sparsity} (a) and (b) show that the sparsity constraint improves the recommendation, but it drops significantly when $\lambda$ is so large that even useful blocks are ignored. Therefore, choosing the balance between real performance and resources becomes crucial for Forward-OFA.

\subsubsection{Influence of structure-based parameters}

Here we take an analysis to figure out how structure-related parameters benefit adaptive networks in our framework. Therefore we ablate some components in our method and compare their performance on three datasets. The ablated models are as follows:
\begin{itemize}
    \item \textbf{w.o. structural vector} does not take the distribution vector(Equation \ref{beta}) as an auxiliary input for the adaptive parameters. This allows the parameters to bypass conflicting gradient updates but ignore the structural information.
    \item \textbf{random block} means we only decides how many blocks will be used, but which blocks are used is random.
    \item \textbf{first $k$ block} uses the first $k$ blocks in the adaptive networks, where $k$ is the number of blocks.
    \item \textbf{last $k$ block} consistently uses the last $k$ blocks in the adaptive networks.
\end{itemize}

The detailed performance is shown in Table \ref{tab:experiment_ablation}, from which we have the following observations: \textbf{1)} Parameters containing the structural information perform better than those without on all datasets, proving the necessity of the structure we generated as input. \textbf{2)} Not using model-specified blocks will yield poor results, whether using the first few or the last few blocks. We believe that not only the exact number of blocks, but the combination of blocks is crucial. Therefore, only by considering both the structural parameters and the adaptive structure can the model that best suits the interests of the device be obtained.

\begin{table}[htb]
\caption{Analysis of Structural Parameters} 
\vspace{-0.3cm}
\label{tab:experiment_ablation}
\centering
\resizebox{1.0\linewidth}{!}{
    \begin{tabular}{c|c|c|c|c|c|c|c}
    \toprule[2pt]

     \multirow{3}{*}{\textbf{Model}} & \multirow{3}{*}{\textbf{Method}} &  \multicolumn{6}{c}{\textbf{Dataset}} \\
     \cline{3-8}
     & & \multicolumn{2}{c|}{\textbf{MovieLens-1M}} & \multicolumn{2}{c|}{\textbf{MovieLens-10M}} & \multicolumn{2}{c}{\textbf{Amazon-Game}} \\
     \cline{3-8}
     & & \textbf{NDCG} & \textbf{Hit} & \textbf{NDCG} & \textbf{Hit} & \textbf{NDCG} & \textbf{Hit} \\
     \bottomrule[1pt]
     \bottomrule[1pt]
     
     \multirow{6}{*}{\textbf{SASRec}} & \texttt{DeviceRec} & 0.0969 & 0.1816 & 0.0718 & 0.1308 & 0.0359 & 0.0541  \\
     & \texttt{random block} & 0.0714 & 0.1325 & 0.0564 & 0.1045 & 0.0462 & 0.0696  \\
     & \texttt{first $k$ block} & 0.0174 & 0.0371 & 0.0030 & 0.0067 & 0.0384 & 0.0546  \\
     & \texttt{last $k$ block} & 0.0699 & 0.1285 & 0.0836 & 0.1464 & 0.0481 & 0.733  \\
     & \texttt{w.o. structural vector} & 0.1166 & 0.2068 & 0.0872 & 0.1504 & 0.0472 & 0.0714  \\
     \rowcolor{gray!40}
     \cellcolor{white} & \texttt{Forward-OFA} & \textbf{0.1182} & \textbf{0.2081} & \textbf{0.0905} & \textbf{0.1566} & \textbf{0.0487} & \textbf{0.0754}  \\

     \midrule[1pt]
     \midrule[1pt]

    \multirow{6}{*}{\textbf{NextItNet}} & \texttt{DeviceRec} & 0.0975 & 0.1846 & 0.0638 & 0.1166 & 0.0275 & 0.0425 \\
    & \texttt{random block} & 0.0346 & 0.0644 & 0.0306 & 0.0571 & 0.0343 & 0.0468  \\
     & \texttt{first $k$ block} & 0.0969 & 0.1795 & 0.0812 & 0.1425 & 0.0455 & 0.0663  \\
     & \texttt{last $k$ block} & 0.0082 & 0.0187 & 0.0132 & 0.0294 & 0.0184 & 0.0214  \\
     & \texttt{w.o. structural vector} & 0.1209 & 0.2141 & 0.0812 & 0.1412 & 0.0456 & 0.0662  \\
     \rowcolor{gray!40}
     \cellcolor{white} & \texttt{Forward-OFA} & \textbf{0.1226} & \textbf{0.2141} & \textbf{0.0824} & \textbf{0.1429} & \textbf{0.0477} & \textbf{0.0708}  \\

    \bottomrule[2pt]
    \end{tabular}
}
\end{table}

\subsubsection{Comparision between Forward-OFA and smaller baselines}
As discussed above, our framework can obtain better performance with smaller models. Inspired by this, we begin to question whether this improvement is simply due to using small models rather than adaptive networks. To further clarify the necessity of adaptive networks, we replace all the baselines in Table \ref{tab:experiment_rec} with smaller models, each of which is $\frac{1}{3}$ the size of the backbones in Table \ref{tab:experiment_rec}. We conduct experiments on three datasets and list the results in Table \ref{tab:experiment_rec_small}. The first observation is that larger models may not always provide better performance. While most larger models achieve better results, on some datasets, the smaller SASRec performs better, which is consistent with what we introduced in Figure \ref{intro}. Another observation is that despite smaller models used by other baselines, our method still achieves the best performance, especially on Movielens-1M and Movielens-10M where our approach has improved by approximately 20$\%$ to 30$\%$. It makes sense that users' latent interests can only be well modeled with adaptive models instead of uniformly large or small models.

\begin{table}[htb]
  \caption{Comparison between our model and others based on small models. We use a two-layer SASRec and a four-layer NextitNet for each baseline in this experiment. $\ast$ denotes the baseline uses small models compared to those in Table \ref{tab:experiment_rec}}
\vspace{-0.3cm}
\label{tab:experiment_rec_small}
\centering
\resizebox{1.0\linewidth}{!}{
    \begin{tabular}{c|c|c|c|c|c|c|c}
    \toprule[2pt]

     \multirow{3}{*}{\textbf{Model}} & \multirow{3}{*}{\textbf{Method}} &  \multicolumn{6}{c}{\textbf{Dataset}} \\
     \cline{3-8}
     & & \multicolumn{2}{c|}{\textbf{MovieLens-1M}} & \multicolumn{2}{c|}{\textbf{MovieLens-10M}} & \multicolumn{2}{c}{\textbf{Amazon-Game}} \\
     \cline{3-8}
     & & \textbf{NDCG} & \textbf{Hit} & \textbf{NDCG} & \textbf{Hit} & \textbf{NDCG} & \textbf{Hit} \\
     \bottomrule[1pt]
     \bottomrule[1pt]
     
     \multirow{8}{*}{\textbf{SASRec}} & \texttt{DeviceRec$^{\ast}$} & 0.0998 & 0.1897 & 0.0733 & 0.1351 & 0.0434 & 0.0656  \\
     & \texttt{finetune$^{\ast}$} & 0.1000 & 0.1902 & 0.0705 & 0.1324 & 0.0430 & 0.0647  \\
     & \texttt{DUET$^{\ast}$} & 0.1128 & 0.2018 & 0.0867 & 0.1499 & 0.0467 & 0.0717  \\
     & \texttt{STTD$^{\ast}$} & 0.0964 & 0.1836 & 0.0722 & 0.1331 & 0.0455 & 0.0694 \\
     & \texttt{Rare Gem$^{\ast}$} & 0.0993 & 0.1877 & 0.0733 & 0.1353 & 0.0450 & 0.0683 \\
     & \texttt{Gater$^{\ast}$} & 0.0991 & 0.1869 & 0.0728 & 0.1350 & 0.0467 & 0.0701 \\
     \cline{2-8}
     \rowcolor{gray!40}
     \cellcolor{white} & \texttt{Forward-OFA} & \textbf{0.1182} & \textbf{0.2081} & \textbf{0.0905} & \textbf{0.1566} & \textbf{0.0487} & \textbf{0.0754}  \\
     \rowcolor{skyblue!70}
     \cellcolor{white} & \texttt{Improv.} & 18.49$\%$ & 9.40$\%$ & 23.67$\%$ & 16.17$\%$ & 12.31$\%$ & 14.84$\%$ \\

     \midrule[1pt]
     \midrule[1pt]

    \multirow{8}{*}{\textbf{NextItNet}} & \texttt{DeviceRec$^{\ast}$} & 0.0947 & 0.1798 & 0.0638 & 0.1166 & 0.0298 & 0.0471  \\
    & \texttt{finetune$^{\ast}$} & 0.0883 & 0.1755 & 0.0610 & 0.1136 & 0.0303 & 0.0474  \\
     & \texttt{DUET$^{\ast}$} & 0.1177 & 0.2081 & 0.0815 & 0.1369 & 0.0455 & 0.664  \\
     & \texttt{STTD$^{\ast}$} & 0.0915 & 0.1753 & 0.0621 & 0.1149 & 0.0262 & 0.0422 \\
     & \texttt{Rare Gem$^{\ast}$} & 0.1025 & 0.1902 & 0.0684 & 0.1243 & 0.0329 & 0.0513\\
     & \texttt{Gater$^{\ast}$} & 0.0973 & 0.1834 & 0.0630 & 0.1153 & 0.0309 & 0.0453 \\
     \cline{2-8}
     \rowcolor{gray!40}
     \cellcolor{white} & \texttt{Forward-OFA} & \textbf{0.1226} & \textbf{0.2141} & \textbf{0.0826} & \textbf{0.1429} & \textbf{0.0477} & \textbf{0.0708}  \\
     \rowcolor{skyblue!70}
     \cellcolor{white} & \texttt{Improv.} & 29.42$\%$ & 19.06$\%$ & 29.05$\%$ & 22.55$\%$ & 59.81$\%$ & 50.38$\%$ \\

    \bottomrule[2pt]
    \end{tabular}
}
\end{table}

\subsection{Complexity and Privacy Analysis(RQ3)}
As \emph{Forward-OFA} takes user historical embeddings as input, uploading the sequence or embeddings to the cloud can lead to privacy issues by potentially leaking the device’s privacy issues. In contrast, we address the privacy problem by placing the structure generator (a GRU and a fully connected layer) and the sequence extractor of the structural mapper(a GRU) on device. At the beginning of each session or when user interests dramatically change\cite{yao2022device,qian2022intelligent}, device would extract its own interest, select network paths itself and send them to cloud, which can protect user privacy because cloud can only get the sequence features extracted by devices. \textbf{Other modules will be saved on cloud to prevent much burden for devices}. Besides, as mentioned in Section \ref{Preliminary} the candidate embeddings of this session will be stored in the local cache, there is no need to communicate with cloud in this session anymore.

For the complexity, we show the parameters and FLOPs of each component in Table \ref{tab: complexity}. \emph{Forward-OFA} only adds a small fraction of parameters($0.012 \times$ of SASRec and $0.005 \times$ of NextItNet), which is tolerable for resource-sensitive devices as \textbf{once deployed, these modules won't be updated from the cloud frequently like parameters}. Moreover, the extra modules on device take much less time to do inference than the original model. The inference of \emph{Forward-OFA} will only happen when interests on device change dramatically or at the beginning of each session, occasional lightweight inference prevents it from occupying a large number of the device's resources for a long time. As for modules on cloud, even if they consume more parameters, they are shared among all devices and \textbf{can be executed in parallel}. Additionally, abundant computing resources on cloud make it efficient to hold these modules.

\begin{table}[htb]
\caption{FLOPs and Param of each component on devices and cloud. Backbone denotes the original sequential recommender, while Forward-device and Forward-cloud denote the extra component included by Forward-OFA, the former is stored on device while the latter is stored on cloud.} 
\vspace{-0.3cm}
\label{tab: complexity}
\centering
\resizebox{1.0\linewidth}{!}{
    \begin{tabular}{c|c|p{2.2cm}<{\centering}|p{2.2cm}<{\centering}|p{2.2cm}<{\centering}}
    \toprule[2pt]
    \multirow{2}{*}{Model} & \multirow{2}{*}{Metric} & \multicolumn{3}{c}{Component} \\
    \cline{3-5}
    & & Backbone & Forward-device & Forward-cloud \\
    \midrule[1pt]
    \multirow{2}{*}{SASRec} & FLOPs & 0.6244 & 0.2544 & 0.2621 \\
    \cline{2-5}
    & Param & 3.9936 & 0.0503 & 8.1120\\
    \midrule[1pt]
    \midrule[1pt]
    \multirow{2}{*}{NextItNet} & FLOPs & 1.5053 &  0.2548 & 0.7864 \\
    \cline{2-5}
    & Param & 9.5846 & 0.0507 & 19.2691 \\

    \bottomrule[2pt]
    \end{tabular}
}
\vspace{-0.3cm}
\end{table}

\subsection{Case Study(RQ4)}

\begin{figure}[htb]
  \centering
  \includegraphics[width=\linewidth]{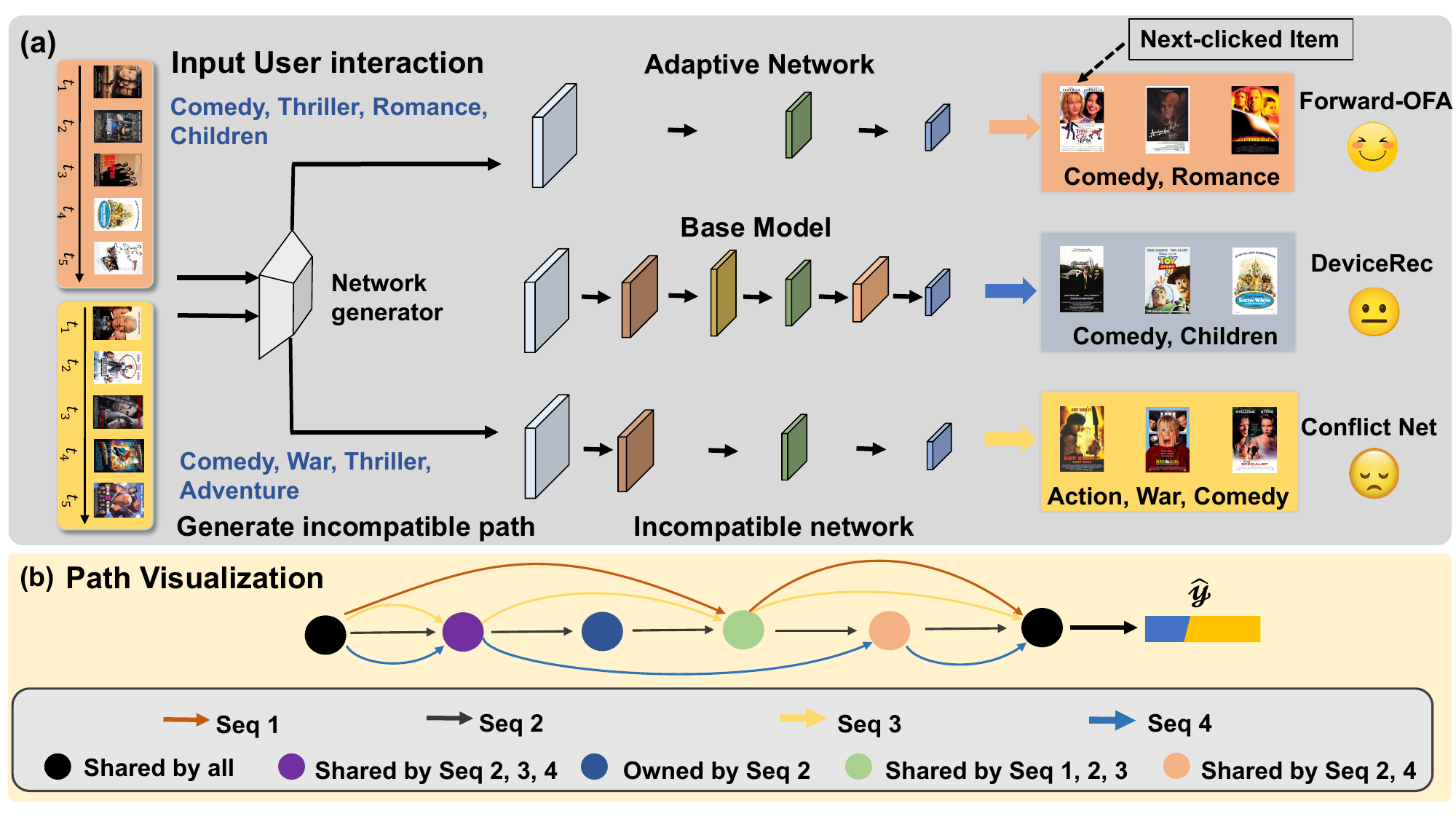}
  \vspace{-0.5cm}
  \caption{(a) The influence of compatible networks for Forward-OFA. The compatible networks(orange one) obtain the best performance compared to the trained model or conflict network generated with another sequence(yellow one). (b) Visualization of assigned paths of these interactions.}
  \label{fig:case_study}
  \vspace{-0.3cm}
\end{figure}

To better understand the importance of adaptive networks for recommendation, we present the case study results from the Movielens-10M dataset. Specifically, we sample four individual interactions and generate the corresponding networks through \emph{Forward-OFA} as shown in Figure \ref{fig:case_study}. The adaptive path(structure) is presented in Figure \ref{fig:case_study} (b), where we can observe that different user interests correspond to different paths. Apart from this, some sequences(1, 3, 4) only require part of the blocks to finish inference, while sequence 2 needs all blocks. Without compatible networks, the final recommendation list will be a deviation from expected results.

For a clearer understanding, we use the \textcolor{orange}{orange} sequence(denoted as \emph{P}) as the input of recommender and consider the following three networks: (1) the adaptive network generated from \emph{Forward-OFA}. (2) the dense model trained with data on cloud. (3) incompatible networks generated using \emph{Forward-OFA} and \textcolor{yellow}{yellow} interactions(denoted as \emph{Y}). Figure \ref{fig:case_study}(a) demonstrates that the adaptive network has the least parameters and provides the best result(with the next-predicted item and relevant items). The trained network gives moderate results, which contain relevant items but not the ones that users are most interested in. Conversely, when utilizing \emph{Y} to generate masks, which exhibit disparate preferences compared to \emph{P}, the outcomes are unfavorable, yielding few relevant movie recommendations. These results further demonstrate that only networks learned by \emph{Forward-OFA} can best fit user requirements.

\section{CONCLUSION}
In this paper, we propose an efficient framework, Forward-OFA to construct local adaptive networks in only one forward pass. The compatible model provides better performance and lowers computational costs, as only a small fraction of blocks are necessary and beneficial for each device. Extensive experiments and comprehensive analysis of various real-world datasets and widely used sequential recommenders demonstrate the feasibility and effectiveness of the Forward-OFA. Furthermore, this work can be viewed as an initiative to explore the possibility of searching for compatible and lightweight networks for specific tasks.

\section*{Acknowledgement}
This work was supported by the 2030 National Science and Technology Major Project (2022ZD0119100), the National Natural Science Foundation of China (62376243, 62402429, 62441605, 62037001), the Starry Night Science Fund at Shanghai Institute for Advanced Study (Zhejiang University), the Key Research and Development Program of Zhejiang Province(2024C03270), ZJU Kunpeng\&Ascend Center of Excellence. This work was also supported by Ant group. The author gratefully acknowledges the support of Zhejiang University Education Foundation Qizhen Scholar Foundation.

\bibliographystyle{ACM-Reference-Format}
\balance
\bibliography{sample-base}

\appendix

\section{Experimental Setup}

\subsection{Datasets}
\label{preprocess}
We evaluate our framework on four public benchmarks in recommender system\cite{harper2015movielens,lakkaraju2013s} \emph{Movielens-1M}\footnote{\url{https://grouplens.org/datasets/movielens/1m}}, \emph{Movielens-10M}\footnote{\url{https://grouplens.org/datasets/movielens/10m}}, \emph{Amazon-Food}, and \emph{Amazon-Game}\footnote{\url{https://nijianmo.github.io/amazon/index.html}}. The statistics of which are presented in table \ref{dataset}. Consistent with prior research, all user-item pairs with positive ratings in the dataset are considered positive samples. Owing to variations in sparsity across the datasets, we exclude users and items with fewer than 20 interactions to provide reliable results. To construct sequential scenarios of user interaction in practical applications, we order each user's interactions by their interaction time. The last interaction of each user will be used for test and others will be used for training.

\begin{table}[htb]
  \caption{Statistics of the datasets.}
  \label{dataset}
  \centering
  \resizebox{1.0\columnwidth}{!}{
  \begin{tabular}{lccccc}
    \toprule
    Dataset     & \#Users     & \#Items     & \#Interactions     & \#SeqLen     & \#Sparsity \\
    \midrule
    ML-1M & 6,040  & 3,012  & 994,852  & 164.71  & 94.52\%     \\
    ML-10M & 69,878  & 8,882  & 9,983,342  & 142.87  & 98.39\%     \\
    Food     & 27,149  & 51,885  & 517,548  & 19.06  & 99.97\%     \\
    Game\    & 11,292  & 68,308  & 223,696  & 19.81  & 99.91\%     \\
    \bottomrule
  \end{tabular}
  }
\end{table}

\subsection{Baseline}

\textbf{Base Models.} Given that current sequential recommenders predominantly consist of Transformer and CNN, we choose \emph{SASRec}\cite{kang2018self} and \emph{NextItNet}\cite{yuan2019simple} as the base model, upon which we incorporate different methods:
\begin{itemize}
    \item \textbf{SASRec}\cite{kang2018self} integrates a self-attention mechanism \cite{vaswani2017attention} into recommender system, thereby enhancing the capture of a user’s interests by considering the individual impact of each item in the historical sequence on the target item.
    \item \textbf{NextItNet}\cite{yuan2019simple} employs a convolutional neural network to introduce local perception and parameter sharing. It stands as the pioneering work to utilize residual learning in recommender system, effectively modeling long-range and intricate dependencies within historical sequences.
\end{itemize}

\begin{algorithm}[htb]
\begin{flushleft}
\caption{Forward-OFA on Cloud}
    
\textbf{Module 1:}~\colorbox{gray!30}{$\rhd$~\emph{Structure-related Parameters Mapper}}
    \resizebox{0.46\textwidth}{!}{
    \begin{tcolorbox}[sharp corners, colframe=gray!80!white, colback=white, boxrule=0.5mm, left=0pt, right=0pt, top=-3pt, bottom=-4pt, boxsep=5pt]
    \begin{algorithmic}
    \State \textbf{Target}: Latent Interests $h$, $L$ Parameter Mappers $\{H^{\prime}_k\}_{k=1,..., L}$ $\mapsto$ Structure-related Parameters $W$
    \Statex \Comment{Excuted in parallel}
    \State \textbf{Input}: Latent Interests $h$, $L$ Parameter Mappers $\{H^{\prime}_k\}_{k=1,..., L}$
    \State \textbf{Output}: Structure-related Parameters $W$
    \end{algorithmic}
    \end{tcolorbox}
    }

\textbf{Module 2:}~\colorbox{gray!30}{$\rhd$~\emph{Model Assembler}}
    \resizebox{0.46\textwidth}{!}{
    \begin{tcolorbox}[sharp corners, colframe=gray!80!white, colback=white, boxrule=0.5mm, left=0pt, right=0pt, top=-3pt, bottom=-4pt, boxsep=5pt]
    \begin{algorithmic}
    \State \textbf{Target}: Structure Representation $\{\beta_{i,0}, \beta_{i,1}\}$, Structure-related Parameters $W$ $\mapsto$ Device-specific Networks $M_d$ 
    \State \textbf{Input}: Structure Representation $\{\beta_{i,0}, \beta_{i,1}\}$, Structure-related Parameters $W$
    \State \textbf{Output}: Device-specific Networks $M_d$
    \end{algorithmic}
    \end{tcolorbox}
    }

\label{alg:cloud}
\end{flushleft}
\end{algorithm}

\begin{algorithm}[htb]
\begin{flushleft}
\caption{Forward-OFA on Device}
    
\textbf{Module 1:}~\colorbox{gray!30}{$\rhd$~\emph{Structure Controller}}
    \resizebox{0.46\textwidth}{!}{
    \begin{tcolorbox}[sharp corners, colframe=gray!80!white, colback=white, boxrule=0.5mm, left=0pt, right=0pt, top=-3pt, bottom=-4pt, boxsep=5pt]
    \begin{algorithmic}
    \State \textbf{Target}: Real Time Interaction $X_d$, Sequence Extractor $E$, Structure Controller $H$ $\mapsto$ Structure Representation $\{\beta_{i,0}, \beta_{i,1}\}$
    \State \textbf{Input}: Real Time Interaction $X_d$, Sequence Extractor $E$, Structure Controller $H$
    \State \textbf{Output}: Structure Representation $\{\beta_{i,0}, \beta_{i,1}\}$
    \end{algorithmic}
    \end{tcolorbox}
    }

\textbf{Module 2:}~\colorbox{gray!30}{$\rhd$~\emph{Privacy-focused Interest Extraction}}
\resizebox{0.46\textwidth}{!}{
    \begin{tcolorbox}[sharp corners, colframe=gray!80!white, colback=white, boxrule=0.5mm, left=0pt, right=0pt, top=-3pt, bottom=-4pt, boxsep=5pt]
    \begin{algorithmic}
    \State \textbf{Target}: Real Time Interaction $X_d$, Structure Representation $\{\beta_{i,0}, \beta_{i,1}\}$, Sequence Extractor $E^{\prime}$ $\mapsto$  Latent Interests $h$
    \State \textbf{Input}: Real Time Interaction $X_d$, Structure Representation $\{\beta_{i,0}, \beta_{i,1}\}$, Sequence Extractor $E^{\prime}$
    \State \textbf{Output}: Latent Interests $h$
    \end{algorithmic}
    \end{tcolorbox}
    }

\textbf{Module 3:}~\colorbox{gray!30}{$\rhd$~\emph{Compatible Network Requests}}
\resizebox{0.46\textwidth}{!}{
    \begin{tcolorbox}[sharp corners, colframe=gray!80!white, colback=white, boxrule=0.5mm, left=0pt, right=0pt, top=-3pt, bottom=-4pt, boxsep=5pt]
    \begin{algorithmic}
    \State \textbf{Target}: Structure Representation $\{\beta_{i,0}, \beta_{i,1}\}$, Latent Interests $h$ $\mapsto$ Device-specific Networks $M_d$
    \Statex \Comment{Send requests to cloud.}
    \State \textbf{Input}: Structure Representation $\{\beta_{i,0}, \beta_{i,1}\}$, Latent Interests $h$
    \State \textbf{Output}: Device-specific Networks $M_d$
    \end{algorithmic}
    \end{tcolorbox}
    }

\textbf{Module 4:}~\colorbox{gray!30}{$\rhd$~\emph{On-device Recommendation}}
\resizebox{0.46\textwidth}{!}{
    \begin{tcolorbox}[sharp corners, colframe=gray!80!white, colback=white, boxrule=0.5mm, left=0pt, right=0pt, top=-3pt, bottom=-4pt, boxsep=5pt]
    \begin{algorithmic}
    \State \textbf{Target}: Real Time Interaction $X_d$, Device-specific Networks $M_d$ $\mapsto$ Next-clicked Item $y_d^{t^\prime}$
    \State \textbf{Input}: Real Time Interaction $X_d$, Device-specific Networks $M_d$
    \State \textbf{Output}: Next-clicked Item $y_d^{t^\prime}$
    \end{algorithmic}
    \end{tcolorbox}
    }
\label{alg:device}
\end{flushleft}
\end{algorithm}

\noindent \textbf{Base Methods.}
\begin{itemize}
    \item \textbf{Device-Rec} trains a unified network on cloud and directly sends it to each device. Once deployed, these networks are not updated any further.
    \item \textbf{Finetune} initially trains a model with all available data on cloud. Each device subsequently fine-tunes its model with local data to better align with its interests.
    \item \textbf{DUET}\cite{lv2023duet} requests different parameters from cloud for each device, thereby achieving significant progress in device-cloud collaborative learning.
    \item \textbf{Rare Gem}\cite{sreenivasan2022rare} proposes identifying less important elements within a given network and subsequently training the sparse network to achieve comparable performance while reducing model size.
    \item \textbf{STTD}\cite{xia2022device} represents each layer with small matrices through multiplication to reduce parameters. However, additional computation is required to reconstruct the original layers.
    \item \textbf{Gater}\cite{chen2019you} employs structural pruning to remove redundant filters, thereby enhancing the efficiency of inference.
\end{itemize}

\section{Pseudocode of Forward-OFA}
\label{Pseudocode}
We briefly introduce Forward-OFA on cloud and device in Algorithm \ref{alg:cloud} and \ref{alg:device}. First, as shown in Algorithm \ref{alg:device}, the compatible network path $\{\beta_{i,0}, \beta_{i,1}\}$ is calculated locally based on the real-time interaction, and the path is then used to encode latent interests $h$ to protect privacy. This process is both memory and calculation-efficient and does not occupy much resource on device. Then device will send its latent interest to cloud to finish the network assembly. In Algorithm \ref{alg:cloud}, parameter mapping for each layer on cloud can be executed in parallel. Compatible networks will be sent back to finish later recommendations.

\section{Association between Neural Architecture Search and Forward-OFA}
\noindent \textbf{Smilarity.} Both NAS (Neural Architecture Search) and Forward-OFA aim to identify suitable networks for devices that adapt to specific data distributions while minimizing resource consumption. They both address an often overlooked research problem: the fundamental significance of network structures to various data distributions. Forward-OFA is also partially motivated by those methods to detect light subnetworks for each device.
 
\noindent \textbf{Difference.} However, NAS-based methods\cite{zhao2021few, zoph2016neural, chen2022autogsr} have to search appropriate structures in advance and retrain local data to get device-specific networks. The whole subnet-constructing process takes a long time, making it impractical for billions of users.
For example, the recent method LitePred\cite{feng2024litepred} matches and finetunes models with device data, requiring substantial resources and leading to longer responses. Besides, limited on-device interactions may cause overfitting or suboptimal performance. In contrast, Forward-OFA achieves adaptation through a single forward pass, directly mapping local interests to networks. We also conducted an additional comparable experiment on LitePred, where the results with ours are shown in Table \ref{tab: NAS}. Owing to the accurate matching process, \emph{LightPred} successfully outperforms \emph{DeviceRec} and \emph{Finetune}, demonstrating the necessity of discovering valuable subnetworks. However, as mentioned above, limited on-device interactions and various latent device interests restrict its potential to learn compatible parameters, leading to a large performance drop compared to Forward-OFA which directly builds networks from real-time interactions.

\begin{table}[htb]
  \caption{Comparation between NAS-based method LitePred and Forward-OFA.} 
\vspace{-0.2cm}
\label{tab: NAS}
\centering
\resizebox{1.0\linewidth}{!}{
    \begin{tabular}{c|c|c|c|c|c|c|c}
    \toprule[2pt]

     \multirow{3}{*}{\textbf{Model}} & \multirow{3}{*}{\textbf{Method}} & \multicolumn{6}{c}{\textbf{Dataset}} \\
     \cline{3-8}
     & & \multicolumn{2}{c|}{\textbf{MovieLens-1M}} & \multicolumn{2}{c|}{\textbf{MovieLens-10M}} & \multicolumn{2}{c|}{\textbf{Amazon-Game}} \\
     \cline{3-8}
     & & \textbf{NDCG $\uparrow$} & \textbf{Hit $\uparrow$} & \textbf{NDCG $\uparrow$} & \textbf{Hit $\uparrow$} & \textbf{NDCG $\uparrow$} & \textbf{Hit $\uparrow$} \\
     \bottomrule[1pt]
     \bottomrule[1pt]

     \multirow{4}{*}{\textbf{SASRec}} & \texttt{DeviceRec} & 0.0969 & 0.1816 & 0.0718 & 0.1308 & 0.0359 & 0.0541\\
     & \texttt{Finetune} & 0.0939 & 0.1793 & 0.0681 & 0.1272 & 0.0343 & 0.0537\\
     & \texttt{LitePred} & 0.0990 & 0.1877 & 0.0735 & 0.1347 & 0.0419 & 0.0635\\
     \cline{2-8}
     \rowcolor{gray!40}
     \cellcolor{white} & \texttt{Forward-OFA} & \textbf{0.1182} & \textbf{0.2081} & \textbf{0.0905} & \textbf{0.1566} & \textbf{0.0487} & \textbf{0.0754}\\
    
    \midrule[1pt]
    \midrule[1pt]

    \multirow{4}{*}{\textbf{NextItNet}} & \texttt{DeviceRec} & 0.0975 & 0.1846 & 0.0638 & 0.1166 &  0.0275 & 0.0440 \\
    & \texttt{Finetune} & 0.0879 & 0.1715 & 0.0635 & 0.1165 & 0.0279 & 0.0422 \\
    & \texttt{LitePred} & 0.0993 & 0.1828 & 0.0652 & 0.1184 & 0.0310 & 0.0484 \\
    \cline{2-8}
     \rowcolor{gray!40}
     \cellcolor{white} & \texttt{Forward-OFA} & \textbf{0.1226} & \textbf{0.2140} & \textbf{0.0826} & \textbf{0.1429} & \textbf{0.0477} & \textbf{0.0708}\\

    \bottomrule[2pt]
    \end{tabular}
}

\end{table}

\section{Adaptation Time to Changing Interests}
To build insights into the efficient adaptation of Forward-OFA, in this section, we compare both the adaptation time once interests shift and the communication time for the model update.
In this section, we will further validate the efficiency of Forward-OFA in quickly adapting to changing user interests. As shown in Table 8, we found that on-cloud fine-tuning on an Nvidia RTX 3090 GPU(35.58 TFLOPS) takes 1000 times longer than Forward-OFA. For mobile devices like the iPhone 16(1789.4 GFLOPS), on-device fine-tuning can exceed 94 seconds, which is about 10,000 times slower than Forward-OFA.
\begin{table}
\vspace{-0.3cm}
\caption{Adaptation time of Forward-OFA and On-device/cloud Finetune.}
\vspace{-0.3cm}
\label{tab:adaptation}
\centering
\resizebox{1.0\linewidth}{!}{
    \begin{tabular}{c|c|c|c}
    \toprule[2pt]
    \textbf{Model} & \textbf{Forward-OFA} & \textbf{On-cloud Finetune} & \textbf{On-device Finetune} \\
    \cline{1-4}
    SASRec & 0.0011s & 4.6259s & $\ge 94.1878$s \\
    NextItNet & 0.0051s & 4.8764 & $\ge 99.2882$s \\
    \bottomrule[2pt]
    \end{tabular}
}
\end{table}

\begin{figure}[htb]
  \centering
  \vspace{-0.1cm}
  \includegraphics[width=0.6\linewidth]{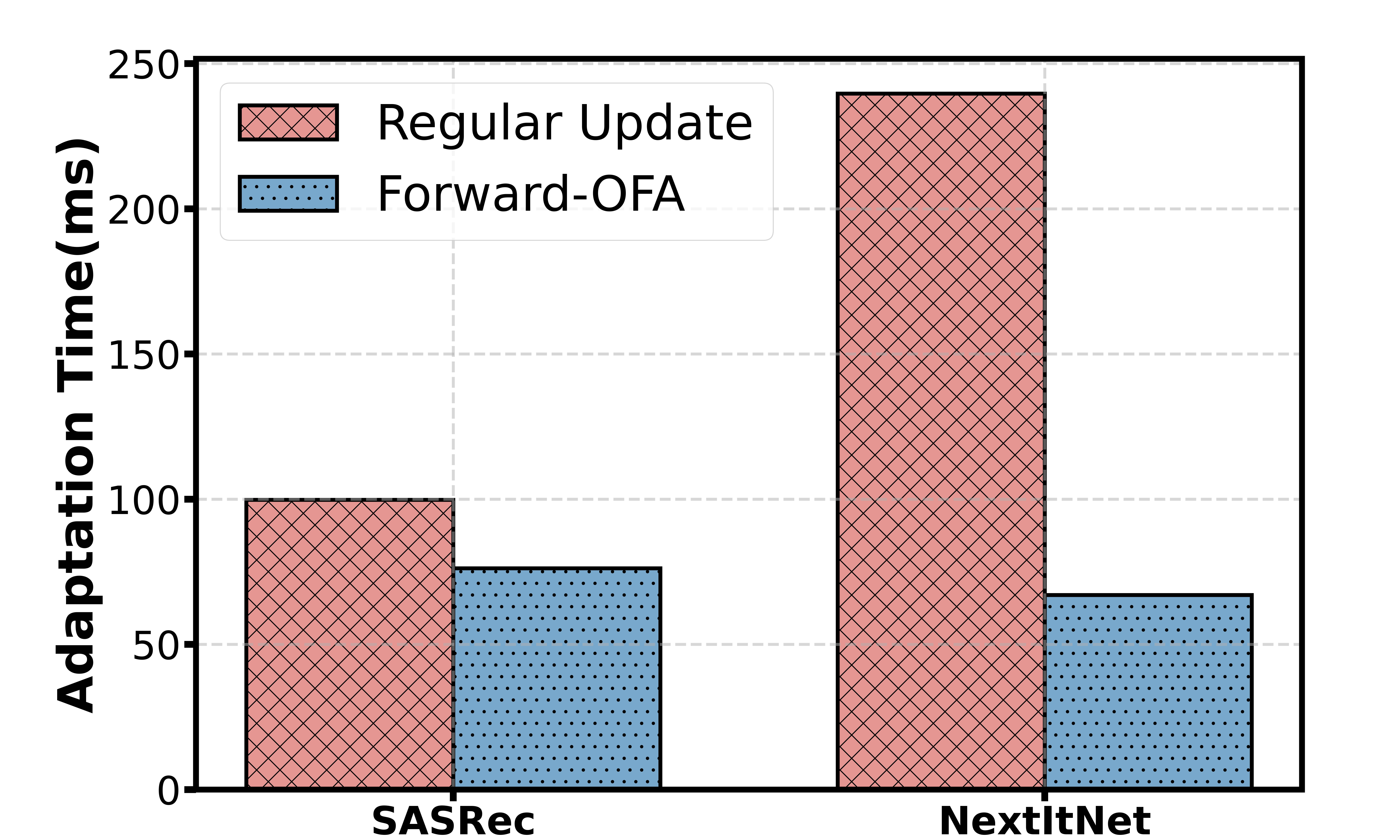}
  \vspace{-0.3cm}
  \caption{Communication time for the model update of Forward-OFA and Regular Strategy.}
  \label{fig: com}
  \vspace{-0.4cm}
\end{figure}

Moreover, Forward-OFA requires significantly less bandwidth as only necessary components are transmitted compared to those enumorous blocks in the original backbones. This results in quicker responses(4G network with 40Mbps) as shown in Figure \ref{fig: com}.

\end{document}